\documentclass[12pt]{article}
\usepackage{axodraw}
\usepackage{graphics}

\newcommand\as{\alpha_{\mathrm{S}}}
\newcommand\smfrac[2]{{\textstyle\frac{#1}{#2}}}
\renewcommand\d{\mathrm{d}}

\newcommand\beeq{\begin{eqnarray}}
\newcommand\eeeq{\end{eqnarray}}
\newcommand\beq{\begin{equation}}
\newcommand\eeq{\end{equation}}
\newcommand\cO{\mathcal{O}}

\begin{document}
\renewcommand{\thefootnote}{\fnsymbol{footnote}}

\begin{titlepage}
\begin{flushright}
  RAL--TR--1999--038 \\ CERN--TH/99-132 \\ hep-ph/9905424
\end{flushright}
\par \vspace{10mm}
\begin{center}
{\Large \bf \boldmath
Corrections of $\cO(\as^2)$ to the\\[1ex]
Forward--Backward Asymmetry~\footnote{This work was 
supported in part 
by the EU Fourth Framework Programme ``Training and Mobility of Researchers'', 
Network ``Quantum Chromodynamics and the Deep Structure of
Elementary Particles'', contract FMRX--CT98--0194 (DG 12 -- MIHT).}}
\end{center}
\par \vspace{2mm}
\begin{center}
{\bf Stefano Catani}~\footnote{On leave of absence from INFN,
Sezione di Firenze, Florence, Italy.}\\
\vspace{5mm}
{Theory Division, CERN}\\
{CH-1211 Geneva 23, Switzerland}
\vspace{1cm}\\
{\bf Michael H. Seymour}\\
\vspace{5mm}
{Rutherford Appleton Laboratory, Chilton}\\
{Didcot, Oxfordshire,  OX11 0QX,  England}
\end{center}
\par \vspace{2mm}
\begin{center} {\large \bf Abstract} \end{center}
\begin{quote}
  \pretolerance=1000
  We calculate the second-order QCD corrections to the forward--backward
  asymmetry in $e^+e^-$ annihilation.  Using the quark axis definition,
  we do not agree with either existing calculation, but the difference
  relative to one of them is small and understood. In particular, we
  point out that the forward--backward asymmetry of massive quarks is
  enhanced by logarithms of the quark mass. This implies that the
  forward--backward asymmetry of massless quarks is not computable in
  QCD perturbation theory and affected by non-power-suppressed
  corrections coming from the non-perturbative fragmentation functions.
  We also calculate the second-order corrections using the
  experimentally-preferred thrust axis definition for the first time.
\end{quote}
\vspace*{\fill}
\begin{flushleft}
  RAL--TR--1999-038 \\ CERN--TH/99-132 \\ May 1999
\end{flushleft}
\end{titlepage}

\setcounter{footnote}{0}
\section{Introduction}
\label{intro}

Some of the most precise determinations of the weak mixing angle
$\sin^2\theta_{ef\!f}$ come from measurements of asymmetries in fermion
production on the Z peak~[\ref{electroweak}].  In particular, the
forward--backward asymmetry of $b$ quarks is measured with a precision of
about 2\%, allowing an extraction of $\sin^2\theta_{ef\!f}$ with almost
per mille accuracy.  However, since we are dealing with quarks in the
final state, we must ensure that QCD corrections, both perturbative and
non-perturbative, are understood to at least the same precision.  From
simple power counting, it is clear that this necessitates including
$\cO(\as^2)$ perturbative and $1/Q$ non-perturbative effects.  Even
these will probably not be enough in the future, when linear $e^+e^-$
colliders are hoped to reach a precision of order 0.1\%~[\ref{NLC}].

The $\cO(\as)$ perturbative corrections were first calculated in
Ref.~[\ref{JLZ}] in the massless approximation.  The mass corrections to
this result were first calculated in Refs.~[\ref{arbuzov}] and were
found to be significant $\sim \as m_b/M_Z$.  These calculations used a
slightly different definition of the asymmetry than the experimental
measurements, which use the thrust axis rather than the quark direction.
This difference was rectified in Refs.~[\ref{DLZ},~\ref{lampe}].

To date there have been two $\cO(\as^2)$ calculations, both in the
massless approximation using the quark direction.  The classic
calculation of Altarelli and Lampe~[\ref{AL}] determined the
$\cO(\as^2)$ coefficient numerically and found it to be small.  This
result has been the basis of all the experimental analyses since.
However, the recent analytical calculation by Ravindran and van
Neerven~[\ref{RvN}] obtained a coefficient about four times bigger.
This discrepancy is comparable to the size of the experimental errors
and needs to be resolved before the final electroweak fits to the LEP1
data can be made.  The $\cO(\as^2)$-calculation using the
experimentally-used thrust axis definition, would also be highly desirable.

In this paper we perform a numerical calculation of the $\cO(\as^2)$
corrections to the forward--backward asymmetry, and compare our results
with the existing calculations.  We also calculate for the first time
the corrections using the thrust axis definition rather than the quark
direction.

The paper is set out as follows.  In Sect.~\ref{sec:def} we define the
forward--backward asymmetry and the closely-related left--right
forward--backward asymmetry~[\ref{BLRV}] and recall some features of the
tree-level and $\cO(\as)$ perturbative calculations.  In
Sect.~\ref{sec:NNLO} we discuss the general set-up of the $\cO(\as^2)$
calculation, and divide it into several parts.  We pay particular
attention to the four-$b$ final state, which will turn out to play an
important r\^ole in our calculation.  In Sect.~\ref{sec:calcn} we make
some final remarks on the details of the calculation, before presenting
our results for the $\cO(\as^2)$ coefficients with the two axis
definitions.  We also compare our results with the existing
calculations.  We discuss the impact of our results in
Sect.~\ref{sec:concl}, and try to estimate the remaining theoretical
errors.  We leave some more technical details of the calculation to
Appendices~A and~B.

\section{Definition and perturbative calculation}
\label{sec:def}

The simplest definition of the 
$b$-quark\footnote{Throughout this paper we explicitly consider the
case of the $b$-quark. The results for the charm quark can be simply obtained by
properly replacing the mass and the electroweak couplings of the massive quark.}
forward--backward asymmetry
$A_{FB}$ is 
\beq 
\label{afbdef} 
A_{FB} = \frac{N_F - N_B}{N_F + N_B} \;\;,
\eeq
where $N_F$ and $N_B$ are the number of $b$ quarks observed in the
forward and backward hemispheres, respectively. 

The axis that identifies the forward
direction can be defined in a variety of ways. However,
for the purpose of making $A_{FB}$ computable in QCD perturbation theory,
the axis must be defined in an infrared- and collinear-safe manner. In this
paper we explicitly consider two different definitions: the $b$-quark 
direction, and the thrust axis direction. 
The thrust axis has a two-fold ambiguity: we use the one
that is nearer the $b$-quark direction. In the following, the 
forward--backward asymmetries with respect to the $b$-quark direction
and to the thrust axis direction are denoted by
$A_{FB}^{b}$ and $A_{FB}^{T}$, respectively.

According to the definition in Eq.~(\ref{afbdef}), 
$A_{FB}$ can be expressed in an equivalent
way in terms of the cross section 
\beq
\label{dcs}
\frac{\d\sigma(e^+e^- \to b + X)}{\d x \;\d\!\cos\theta} 
\eeq
for inclusive $b$-quark production, where $x$ is the fraction of the 
electron energy  carried by the 
$b$ quark and $\theta$ is the angle between the electron momentum and
the direction defining the forward hemisphere (both energies and angles are
defined in the centre-of-mass frame). 

Starting from the distribution in 
Eq.~(\ref{dcs}), we can introduce the forward and backward cross sections
$\sigma_F$ and $\sigma_B$:
\begin{equation}
  \sigma_{F} \equiv
  \int_{0}^{1} \d\!\cos\theta \,\int_{0}^{1} \d x \,
  \frac{\d\sigma}{\d x \;\d\!\cos\theta} \;,
  \;\;\;\;
  \sigma_{B} \equiv
  \int_{-1}^{0} \d\!\cos\theta \,\int_{0}^{1} \d x \,
  \frac{\d\sigma}{\d x \;\d\!\cos\theta} \;,
\end{equation}
and the symmetric and antisymmetric cross sections $\sigma_S$ and
$\sigma_A$:
\beq
\label{saxs}
    \sigma_S = \sigma_F + \sigma_B \;,
    \;\;\;\;
    \sigma_A = \sigma_F - \sigma_B \;.
\eeq
We can then write the forward--backward asymmetry as 
\beq
\label{xsratio}
 A_{FB} = \frac{\sigma_A}{\sigma_S} \;\;.
\eeq 

In the perturbative QCD calculation of $\sigma_S$ and $\sigma_A$, we
have to evaluate the corresponding matrix element squared, which is
given by the product $L_{\mu \nu} T^{\mu \nu}$ of the leptonic and hadronic
tensors $L_{\mu \nu}$ and $T^{\mu \nu}$. Then we could perform the
integration over the final-state parton momenta in $T^{\mu \nu}$ and
finally the integration over the scattering angle $\theta$.
Nonetheless, it is more convenient to use a simplified procedure.  We
can indeed avoid having to explicitly integrate over the scattering
angle, by first performing the angular integration of the leptonic
tensor. Doing this, we can compute $\sigma_S$ and $\sigma_A$ by simply
performing the integration over the final-state parton momenta of the
following projections of the hadronic tensor:
\begin{eqnarray}
  \label{tensorS}
  \sigma_S &\propto& -g_{\mu\nu}\;T^{\mu\nu} \;,\\
  \label{tensorA}
  \sigma_A &\propto& i \epsilon_{\mu\nu\lambda\rho}
  \frac{n^{\lambda}Q^\rho}{n\cdot Q}
  \;T^{\mu\nu}\;,
\end{eqnarray}
where $Q^\mu$ is the total incoming momentum and the light-like
($n^2=0$) vector $n^\mu$ identifies the forward direction.

\subsection{Leading order}
\label{sec:lo}

At the leading order (LO) we have to consider the cross sections for the
process $e^+e^- \to b\bar{b}$ at the tree level and thus, the $b$-quark
direction and the thrust direction coincide.  The tree-level cross
sections $\sigma_S^{(0)}$ and $\sigma_A^{(0)}$ are straightforward to
calculate and the result is\footnote{Unless explicitly mentioned, we
  neglect the $b$-quark mass throughout this paper. At LO the dominant
  mass corrections are proportional to $m_b^2/Q^2$ and can be found, for
  instance, in Ref.~[\ref{RvN}].}  \beeq
\label{sslo}
  \sigma_S^{(0)} &=& \frac{4 \pi \alpha^2 N_c}{3Q^2} \left\{
  e_e^2P_ve_b^2+2\frac{(Q^2-M_Z^2)Q^2}{D_Z(Q^2)}(P_ve_ev_e+P_ae_ea_e)e_bv_b
  \right. \nonumber \\
  &~&\;\;\;\;\; \;\;\;\;\;\;\;\;\;\;\left.
  +\frac{Q^4}{D_Z(Q^2)}\left[ (v_e^2+a_e^2) P_v+2P_av_ea_e \right](v_b^2+a_b^2),
  \right\} \;,\\
\label{salo}
  \sigma_A^{(0)} &=& \frac{4 \pi \alpha^2 N_c}{3Q^2} \; \frac{3}{4} \left\{
  2\frac{(Q^2-M_Z^2)Q^2}{D_Z(Q^2)}(P_ve_ea_e+P_ae_ev_e)e_ba_b
  \right. \nonumber \\
  &~&\;\;\;\;\; \;\;\;\;\;\;\;\;\;\;\;\;\left.
  +\frac{Q^4}{D_Z(Q^2)}\left[ 2P_vv_ea_e+P_a(v_e^2+a_e^2)\right] 2v_ba_b  
  \right\} \;,
\eeeq  
with
\begin{eqnarray}
  P_v &=& 1+P_LP_R \;, \\
  P_a &=& P_L+P_R \;,
\end{eqnarray}  
where $P_L$ is the left-hand-polarization of the electron (+1 = fully
left-handed, 0 = unpolarized, --1 = fully right-handed) and $P_R$ is
the right-hand-polarization of the positron (+1 = fully right-handed
and so forth),
\begin{equation}
  D_Z(Q^2) = (Q^2-M_Z^2)^2+(\Gamma_ZM_Z)^2 \;,
\end{equation}
$e_i$ is the electric charge in units of the proton charge
(i.e.~$e_e=-1$) and the electroweak couplings are:
\begin{eqnarray}
  v_i &=& \frac1{2\sin\theta_w\cos\theta_w}(t_{3i}-2e_i\sin^2\theta_w) \;,\\
  a_i &=& \frac1{2\sin\theta_w\cos\theta_w} t_{3i} \;.
\end{eqnarray}
The ratio between Eqs.~(\ref{sslo}) and (\ref{salo}) is insensitive to 
the fine structure constant $\alpha$ and the number of colours $N_c$ and thus,
at LO the forward--backward asymmetry $A_{FB}^{(0)}$,
\beq
\label{AFB01}
A_{FB}^{(0)} = \frac{\sigma_A^{(0)}}{\sigma_S^{(0)}} \;,
\eeq
gives a direct measurement of the electroweak couplings.
In particular, if we are exactly on the resonance, $Q^2=M_Z^2$, and we
neglect the photon contribution, we obtain
\begin{equation}
  \label{AFB03}
  A_{FB}^{(0)} = \frac34\:
  \frac{\mathcal{A}_e+\mathcal{P}}{1+\mathcal{A}_e\mathcal{P}}\mathcal{A}_b \;,
\end{equation}
where
\begin{eqnarray}
  \mathcal{A}_i &=& \frac{2v_ia_i}{v_i^2+a_i^2} \;, \\
  \mathcal{P} = \frac{P_a}{P_v} &=& \frac{P_L+P_R}{1+P_LP_R} \;.
\end{eqnarray}
Finally, for unpolarized beams, we obtain
\begin{equation}
  \label{AFB04}
  A_{FB}^{(0)} = \frac34\mathcal{A}_e\mathcal{A}_b \;.
\end{equation}
This is the form in which the forward--backward asymmetry is most often 
presented.  It is worth
pointing out however that all of our results will be universal
multiplicative corrections\footnote{At $\cO(\as^2)$ there are some
  non-universal corrections, but we do not 
  explicitly compute them (see the
  discussion in Sects.~\ref{sec:sing} and~\ref{sec:triang}).}, so apply
equally well to any of the forms (\ref{AFB01},~\ref{AFB03})
or~(\ref{AFB04}).

Another important variable is the so-called left--right forward--backward
asymmetry~[\ref{BLRV}], 
\beq 
\label{alrfbdef}
A_{LR,FB} = \frac{N_F(\mathcal{P}=+1)-N_F(\mathcal{P}=-1)
  -N_B(\mathcal{P}=+1)+N_B(\mathcal{P}=-1)}
{N_F(\mathcal{P}=+1)+N_F(\mathcal{P}=-1)
  +N_B(\mathcal{P}=+1)+N_B(\mathcal{P}=-1)} \;\;.
\eeq
Its LO expression can be obtained from
Eqs.~(\ref{sslo}) and~(\ref{salo}), and again neglecting the photon
contribution exactly on the Z resonance, it is given by:
\begin{equation}
  A_{LR,FB}^{(0)} = \frac34\mathcal{A}_b \;.
\end{equation}
Our results apply equally well also to this observable.

\subsection{Next-to-leading-order corrections}
\label{sec:nlo}

At next-to-leading order (NLO), we have to consider the one-loop
cross sections $\sigma^{(1);\mbox{\footnotesize one-loop}}$ 
for the two-parton process $e^+e^- \to b {\bar b}$
and the tree-level cross sections 
$\sigma^{(1);\mbox{\footnotesize tree}}$ for the three-parton
process $e^+e^- \to b {\bar b} g$.  We obtain: 
\begin{equation}
\label{ratio1}
  A_{FB}^{(1)} = \frac
  {\sigma_A^{(0)} + \sigma_A^{(1);\mbox{\footnotesize one-loop}} 
  + \sigma_A^{(1);\mbox{\footnotesize tree}}}
  {\sigma_S^{(0)} + \sigma_S^{(1);\mbox{\footnotesize one-loop}} 
  + \sigma_S^{(1);\mbox{\footnotesize tree}}} \;.
\end{equation}
Each of the cross sections at $\cO(\as)$
is separately divergent, so they have to be
regularized in some way before being combined together.  In any
regularization scheme that preserves the helicity conservation of
massless QCD\footnote{Note that the relation (\ref{1loopcor}) is explicitly 
violated for massive quarks.} (for example,
dimensional regularization), we have the property
\begin{equation}
\label{1loopcor}
  \frac{\sigma_A^{(1);\mbox{\footnotesize one-loop}}}{\sigma_A^0}
  =\frac{\sigma_S^{(1);\mbox{\footnotesize one-loop}}}{\sigma_S^0},
\end{equation}
and hence, if we expand the ratio in Eq.~(\ref{ratio1}) up to 
$\mathcal{O}(\as)$, the one-loop
corrections cancel, and we obtain
\begin{equation}
\label{r1tree}
  A_{FB}^{(1)} = \frac{\sigma_A^{(0)}}{\sigma_S^{(0)}}
  \Biggl(1 + \frac{\sigma_A^{(1);\mbox{\footnotesize tree}}}{\sigma_A^{(0)}} -
  \frac{\sigma_S^{(1);\mbox{\footnotesize tree}}}{\sigma_S^{(0)}}
  \Biggr).
\end{equation}
Although $\sigma_A^{(1);\mbox{\footnotesize tree}}$ and 
$\sigma_S^{(1);\mbox{\footnotesize tree}}$ are each separately divergent
in the soft and collinear regions,
the divergences cancel at the integrand level, and the whole thing can
be calculated in the unregularized theory.

At this order, the different definitions of the
forward--backward asymmetry give different results. As already anticipated,
we consider two possible definitions of the forward direction: 
the $b$-quark direction and the thrust axis direction.  

It is straightforward to calculate the NLO corrections in Eq.~(\ref{r1tree})
analytically with either definition.  We obtain:
\begin{eqnarray}
\label{afb1b}
  A_{FB}^{(1);b} &=& A_{FB}^{(0)} \Biggl(1 - \frac34C_F\frac{\as}{\pi}\Biggr) 
  \simeq A_{FB}^{(0)} \left( 1 - 0.318 \;\as \right) \;, \\
\label{afb1t}  
  A_{FB}^{(1);T} &=& A_{FB}^{(0)} \Biggl(1 - \left\{
    \frac74 - 4\ln\frac32 + \frac{\pi^2}6 + \ln^22 - 
     \frac58\ln3 + 2\mathrm{Li}_2(-\frac12)\right\}
   C_F\frac{\as}{\pi}\Biggr) \\
   &\simeq& A_{FB}^{(0)} \Biggl(1 - 0.670 C_F\frac{\as}{\pi}\Biggr)
   \simeq A_{FB}^{(0)} \Biggl(1 - 0.285 \;\as \Biggr).
\end{eqnarray}
The result in Eq.~(\ref{afb1b}) is well known~[\ref{JLZ}].
The analytical result in Eq.~(\ref{afb1t}) agrees with the numerical 
calculation performed in Refs.~[\ref{DLZ},~\ref{lampe}].
The difference between the two definitions is only about
0.4\% for $\as\sim0.12$. 

We remind the reader that the NLO QCD correction to the symmetric cross
section $\sigma_S$ in the massless limit is equal to the correction
to the $e^+e^-$ total cross section, namely
\beq
\sigma_S = \sigma_S^{(0)} \left( 1 + \frac34C_F\frac{\as}{\pi} + \cO(\as^2)
\right) \;.
\eeq
Thus, Eqs.~(\ref{afb1b}) and (\ref{afb1t}) imply the following results
for the antisymmetric cross sections
\beeq
\label{sabnlo}
\! \! \!\!\sigma_A^{b} &=& \sigma_A^{(0)} \left( 1 + \cO(\as^2) \right) \;, \\
\label{satnlo}
\! \! \!\!\sigma_A^{T} &=& \sigma_A^{(0)}\left( 1 - \left\{
    1 - 4\ln\frac32 + \frac{\pi^2}6 + \ln^22 - 
     \frac58\ln3 + 2\mathrm{Li}_2(-\frac12)\right\}
   C_F\frac{\as}{\pi}\Biggr) + \cO(\as^2) \right) \\
\! \! \!\! &\simeq& \sigma_A^{(0)}\left( 1 + 0.034\as + \cO(\as^2) \right) \,.
\eeeq
The vanishing of the $\cO(\as)$-correction to the antisymmetric cross
section $\sigma_A^{b}$ with respect to the $b$-quark axis in the massless case
was first noticed in Ref.~[\ref{JLZ}].

Unlike at LO, the corrections to $A_{FB}^{(1)}$ due to the finite mass of the 
$b$ quark are of $\cO(m_b/Q)$. The mass corrections have been 
computed in analytic form for the $b$-quark direction~[\ref{arbuzov}] and
numerically for the thrust direction~[\ref{DLZ}].

\section{Contributions at next-to-next-to-leading order}
\label{sec:NNLO}

At next-to-next-to-leading order (NNLO) we have to consider the diagrams
of Figs.~\ref{diags1}--\ref{diags5}. 
The single diagram drawn in Fig.~\ref{diags1}b stands for all the one-loop
diagrams with one virtual gluon. Analogously, the diagram in Fig.~\ref{diags3}
stands for all the tree-level diagrams contributing to the $b{\bar b}gg$ final
state, and so forth.
\begin{figure}[b]
\centerline{%
\begin{picture}(144,108)(0,0)
\Text(0,48)[tl]{(a)}
\SetWidth{1}
\DashLine(0,54)(72,54){12}
\SetWidth{2}
\ArrowLine(72,54)(144,108)
\ArrowLine(144,0)(72,54)
\end{picture}
\hfill
\begin{picture}(144,108)(0,0)
\Text(0,48)[tl]{(b)}
\SetWidth{1}
\DashLine(0,54)(72,54){12}
\Gluon(96,72)(96,36){4}{3}
\SetWidth{2}
\Line(72,54)(96,72)\ArrowLine(96,72)(144,108)
\Line(72,54)(96,36)\ArrowLine(144,0)(96,36)
\end{picture}
\hfill
\begin{picture}(144,108)(0,0)
\Text(0,48)[tl]{(c)}
\SetWidth{1}
\DashLine(0,54)(72,54){12}
\Gluon(96,72)(96,36){4}{3}
\Gluon(120,90)(120,18){4}{6}
\SetWidth{2}
\Line(72,54)(120,90)\ArrowLine(120,90)(144,108)
\Line(72,54)(120,18)\ArrowLine(144,0)(120,18)
\end{picture}}
\vspace{2ex}
\centerline{%
\begin{picture}(144,108)(0,0)
\Text(0,48)[tl]{(d)}
\SetWidth{1}
\DashLine(0,54)(72,54){12}
\ArrowLine(72,54)(96,72)\ArrowLine(96,72)(96,36)\ArrowLine(96,36)(72,54)
\Gluon(96,72)(120,90){4}{3}
\Gluon(96,36)(120,18){-4}{3}
\SetWidth{2}
\ArrowLine(144,0)(120,18)
\ArrowLine(120,18)(120,90)
\ArrowLine(120,90)(144,108)
\end{picture}}
\caption{Some of the diagrams contributing to the $b\bar{b}$ final state
  up to $\mathcal{O}(\as^2)$.  The dashed line represents either the
  axial or vector current, the thick line the $b$ and the thin line
  another quark $q$, which must be summed over flavours, including the
  $b$- and $t$-quark contributions.}
\label{diags1}
\end{figure}
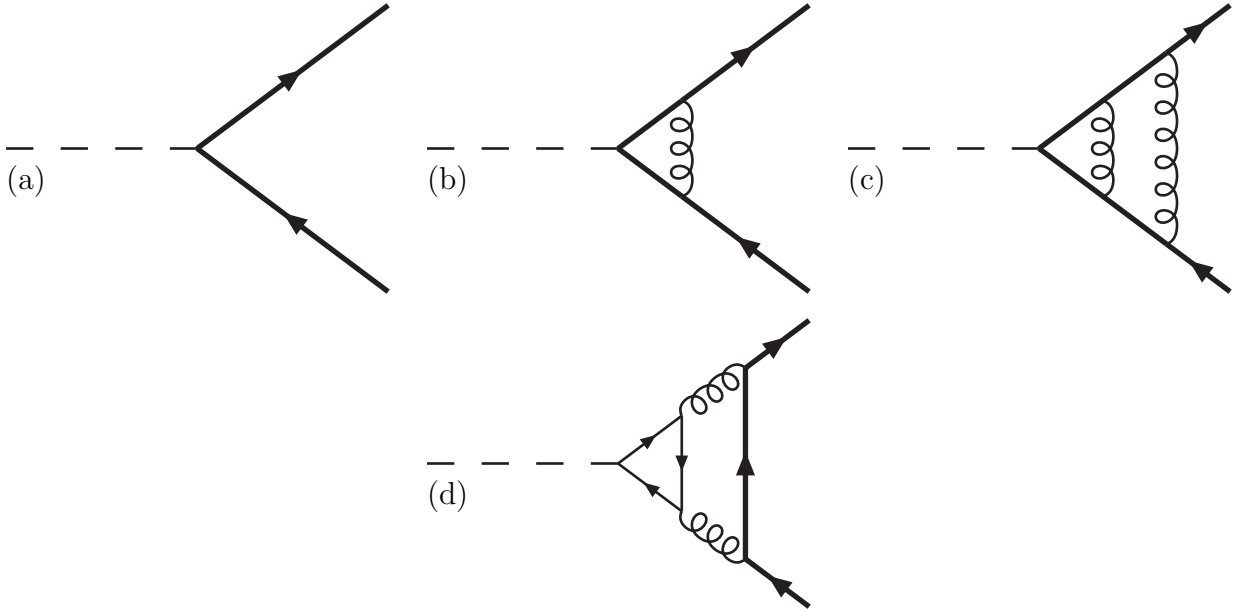
\begin{figure}
\centerline{\hspace{-2em}
\begin{picture}(144,108)(0,0)
\Text(0,48)[tl]{(a)}
\SetWidth{1}
\DashLine(0,54)(72,54){12}
\Gluon(96,72)(144,54){-4}{4}
\SetWidth{2}
\Line(72,54)(96,72)\ArrowLine(96,72)(144,108)
\ArrowLine(144,0)(72,54)
\end{picture}
\hfill
\begin{picture}(144,108)(0,0)
\Text(0,48)[tl]{(b)}
\SetWidth{1}
\DashLine(0,54)(72,54){12}
\GlueArc(72,54)(36,-36.87,36.87){4}{4}
\Gluon(112,54)(144,54){-4}{2}
\SetWidth{2}
\Line(72,54)(96,72)\ArrowLine(96,72)(144,108)
\Line(72,54)(96,36)\ArrowLine(144,0)(96,36)
\end{picture}
\hfill
\begin{picture}(144,108)(0,0)
\Text(0,48)[tl]{(c)}
\SetWidth{1}
\DashLine(0,54)(72,54){12}
\ArrowLine(72,54)(96,72)\ArrowLine(96,72)(96,36)\ArrowLine(96,36)(72,54)
\Gluon(96,72)(144,108){4}{6}
\Gluon(96,36)(120,18){-4}{3}
\SetWidth{2}
\ArrowLine(144,0)(120,18)
\ArrowLine(120,18)(144,36)
\end{picture}\hspace{-2em}}
\caption{Some of the diagrams contributing to the $b\bar{b}g$ final
  state up to $\mathcal{O}(\as^2)$.  The dashed line represents either
  the axial or vector current, the thick line the $b$ and the thin line
  another quark $q$, which must be summed over flavours, including the
  $b$- and $t$-quark~contributions.}
\label{diags2}
\end{figure}
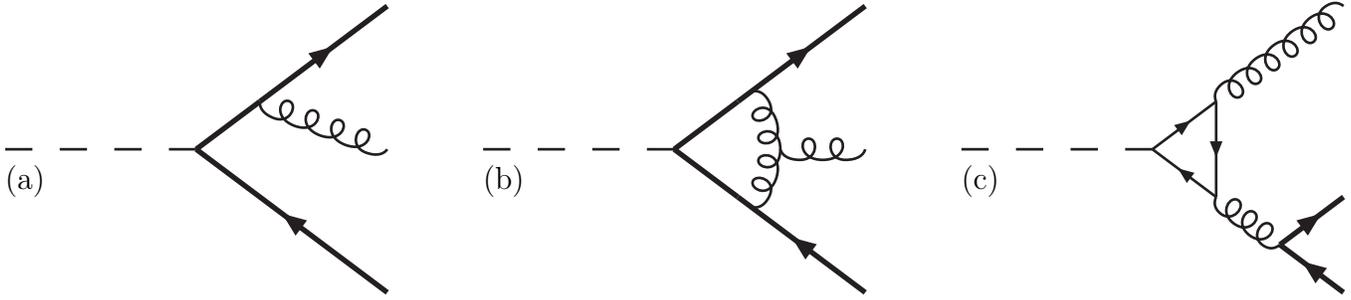
\begin{figure}
\centerline{%
\begin{picture}(144,108)(0,0)
\SetWidth{1}
\DashLine(0,54)(72,54){12}
\Gluon(96,72)(144,66){-4}{4}
\Gluon(96,36)(144,42){4}{4}
\SetWidth{2}
\Line(72,54)(96,72)\ArrowLine(96,72)(144,108)
\Line(72,54)(96,36)\ArrowLine(144,0)(96,36)
\end{picture}}
\caption{One of the diagrams contributing to the $b\bar{b}gg$ final
  state at $\mathcal{O}(\as^2)$.  The dashed line represents either
  the axial or vector current.}
\label{diags3}
\end{figure}
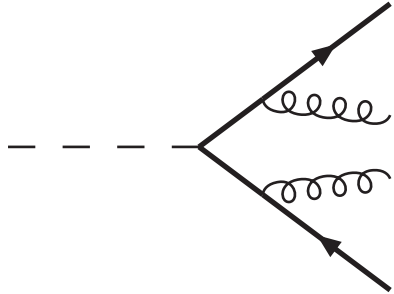
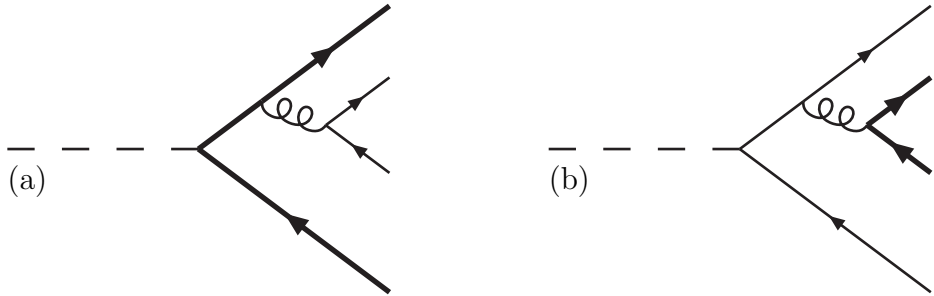
\begin{figure}
\centerline{\hfill
\begin{picture}(144,108)(0,0)
\Text(0,48)[tl]{(a)}
\SetWidth{1}
\DashLine(0,54)(72,54){12}
\Gluon(96,72)(120,63){-4}{2}
\ArrowLine(120,63)(144,81)
\ArrowLine(144,45)(120,63)
\SetWidth{2}
\Line(72,54)(96,72)\ArrowLine(96,72)(144,108)
\ArrowLine(144,0)(72,54)
\end{picture}
\hfill
\begin{picture}(144,108)(0,0)
\Text(0,48)[tl]{(b)}
\SetWidth{1}
\DashLine(0,54)(72,54){12}
\Gluon(96,72)(120,63){-4}{2}
\Line(72,54)(96,72)\ArrowLine(96,72)(144,108)
\ArrowLine(144,0)(72,54)
\SetWidth{2}
\ArrowLine(120,63)(144,81)
\ArrowLine(144,45)(120,63)
\end{picture}\hfill}
\caption{Some of the diagrams contributing to the $b\bar{b}q\bar{q}$ final
  state at $\mathcal{O}(\as^2)$.  The dashed line represents either
  the axial or vector current, the thick line the $b$ and the thin line
  some other quark flavour $q$, with $q \neq b$.}
\label{diags4}
\end{figure}
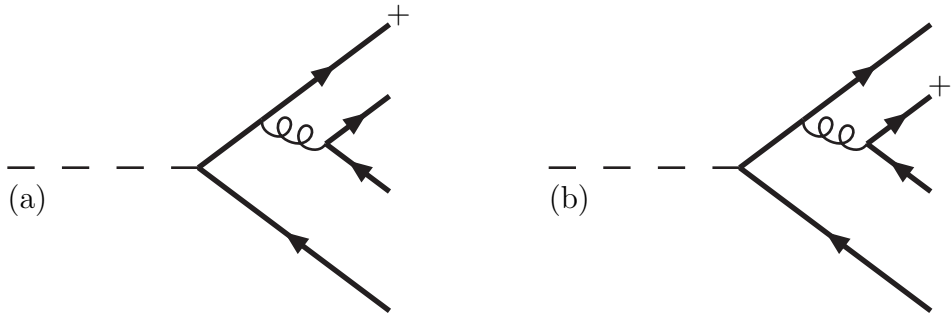
\begin{figure}
\centerline{\hfill
\begin{picture}(144,108)(0,0)
\Text(0,48)[tl]{(a)}
\SetWidth{1}
\DashLine(0,54)(72,54){12}
\Gluon(96,72)(120,63){-4}{2}
\SetWidth{2}
\Line(72,54)(96,72)\ArrowLine(96,72)(144,108)
\ArrowLine(144,0)(72,54)
\ArrowLine(120,63)(144,81)
\ArrowLine(144,45)(120,63)
\Text(148,112)[c]{$+$}
\end{picture}
\hfill
\begin{picture}(144,108)(0,0)
\Text(0,48)[tl]{(b)}
\SetWidth{1}
\DashLine(0,54)(72,54){12}
\Gluon(96,72)(120,63){-4}{2}
\SetWidth{2}
\Line(72,54)(96,72)\ArrowLine(96,72)(144,108)
\ArrowLine(144,0)(72,54)
\ArrowLine(120,63)(144,81)
\ArrowLine(144,45)(120,63)
\Text(148,85)[c]{$+$}
\end{picture}\hfill}
\caption{One of the diagrams contributing to the $b\bar{b}b\bar{b}$ final
  state at $\mathcal{O}(\as^2)$.  The dashed line represents either the
  axial or vector current.  The cross indicates which of the two $b$
  quarks is triggered on: both contributions must be summed.}
\label{diags5}
\end{figure}

We separate the contributions to the cross sections into three classes:
flavour non-singlet $(NS)$, flavour singlet $(S)$, and interference 
(or triangle) $(Tr)$.  
We thus write the cross sections as
\beeq
\label{sdecomp}
\sigma_S &=& \sigma_{S, NS} + \sigma_{S, S}^{(2)} + \sigma_{S, Tr}^{(2)}
+ \cO(\as^3) \;, \\
\label{adecomp}
\sigma_A &=& \sigma_{A, NS} + \sigma_{A, Tr}^{(2)} + \cO(\as^3) \;. 
\eeeq
In this notation, up to $\mathcal{O}(\as)$ there are only non-singlet
contributions. Thus, $\sigma_{S, S}^{(2)}, \sigma_{S, Tr}^{(2)}$ and
$\sigma_{A, Tr}^{(2)}$ are proportional to $\as^2$. Note also that
there are no singlet contributions to the antisymmetric cross section
$\sigma_A$.

The forward--backward asymmetry is decomposed in a similar way. Expanding
the ratio $\sigma_A/\sigma_S$ up to $\cO(\as^2)$, we write
\begin{equation}
\label{afbdec}
  A_{FB}^{(2)} = A_{FB, NS}^{(2)} +
                 \frac{\sigma_A^{(0)}}{\sigma_S^{(0)}}
        \left( \frac{\sigma_{A, Tr}^{(2)}}{\sigma_A^{(0)}}
             - \frac{\sigma_{S, Tr}^{(2)}}{\sigma_S^{(0)}}
             - \frac{\sigma_{S, S}^{(2)}}{\sigma_S^{(0)}} \right) \;\;,
\end{equation}  
where $A_{FB, NS}^{(2)}$ denotes the non-singlet component:
\beq
\label{afbns}
   A_{FB, NS}^{(2)} = \frac{\sigma_{A, NS}}{\sigma_{S, NS}} \;.
\eeq

We now discuss our treatment of each contribution in turn.
The classification of the four-$b$ contribution of Fig.~\ref{diags5}
also warrants additional discussion. 

\vspace{-1ex}
\subsection{Triangle contributions}
\label{sec:triang}

In this class we group all the cross section contributions consisting of
two quark triangles, one attached to each current.  These correspond to
the interference between the diagrams in Figs.~\ref{diags1}d~and~\ref{diags1}a,
between those in Figs.~\ref{diags2}c~and~\ref{diags2}a, and between those 
in Figs.~\ref{diags4}b~and~\ref{diags4}a.  They give
non-universal (i.e.~non-factorizable) corrections to both the symmetric and 
antisymmetric cross
sections.  They are calculated in Ref.~[\ref{AL}] for the $b$-quark axis
definition and found to be very small.  To our knowledge their
contribution to the thrust axis definition has never been calculated,
but we expect it to be similarly small.  We therefore neglect it, 
i.e.~$\sigma_{S, Tr}^{(2)}$ and $\sigma_{A, Tr}^{(2)}$ in Eq.~(\ref{afbdec}), 
from our calculation\footnote{We remind the reader that the triangle
contributions to both $\sigma_S$ and $\sigma_A$ are finite in the massless 
limit~$m_b \to 0$, provided that the sums over quark flavour $q$ in the
diagrams of Figs.~\ref{diags1}d and~\ref{diags2}c run over complete
SU(2)~doublets.}.
 
\vspace{-1ex}
\subsection{Singlet contributions}
\label{sec:sing}

In this class we group the square of the diagrams of Fig.~\ref{diags4}b, 
where the
final-state $b$ quark is not coupled to the current.  In these
contributions the $b$ and $\bar{b}$ are produced in a definite state of
charge conjugation, $C=+1$.  They therefore cannot contribute to the
antisymmetric cross section,~$\sigma_A$.  Their contribution to the
symmetric cross section, $\sigma_S$, is logarithmically enhanced in the
small-mass limit and proportional to $\as^2\ln^3Q^2/m_b^2$.  
An approximate expression for it, denoted by $F^{{\rm Branco}}$, was used in
Ref.~[\ref{AL}].
It is calculated
exactly to $\mathcal{O}(\as^2)$ in Refs.~[\ref{HQ},~\ref{HQresum}], and
the leading and next-to-leading logarithms are summed to all orders in $\as$
in Ref.~[\ref{HQresum}].

Note that the singlet contributions to $\sigma_S$ include an additional term
coming from the $b{\bar b}b{\bar b}$ final state. As discussed in 
Sect.~\ref{sec:4b}, this term is very similar to that described above.
It was missing in the expression denoted by $F^{{\rm Branco}}$ in
Ref.~[\ref{AL}].

In some sense the singlet component is a `background' to the
forward--backward asymmetry measurement and, in fact, in the experimental
analyses (see e.g.~Ref.~[\ref{abbaneo}])
it is statistically subtracted using Monte Carlo event
generators.  We therefore neglect it, 
i.e.\ $\sigma_{S, S}^{(2)}$ in Eq.~(\ref{afbdec}),
from our calculation.

\subsection{Four-\boldmath$b$ contributions}
\label{sec:4b}

The classification of the four-$b$ diagrams of Fig.~\ref{diags5} deserves 
special mention. 
Let us first point out a basic fact. The four-$b$ diagrams of Fig.~\ref{diags5}
contribute to both the $b$-quark cross sections $\sigma_S$ and
$\sigma_A$ and
the $e^+e^-$ total cross section. However, they appear with different
multiplicity factors in the two cases. In the case of the
$e^+e^-$ total cross section the multiplicity factor is simply equal to unity.
In the contribution to the 
\emph{inclusive\/} $b$-quark cross sections $\sigma_S$ and $\sigma_A$, these
diagrams count twice since there are two $b$ quarks in the final state.
This observation is relevant in the discussion that follows and, in particular,
it is important in understanding the results for the
non-singlet component of the symmetric cross section $\sigma_S$
discussed in Sect.~\ref{sec:nonsing}.

After summing and squaring the diagrams in Fig.~\ref{diags5},
we obtain two types of contribution: $i)$~those
that are identical to the contributions of Fig.~\ref{diags4} but with
the other quark $q$ replaced by an untriggered-on $b$ quark, and $ii)$~those 
that are genuine interference terms arising from the fact that the two 
antiquarks are indistinguishable, called the $E$-term in Ref.~[\ref{ERT}].  
The squared diagrams of type~$i)$ are treated as those of Fig.~\ref{diags4},
that is, we lump them together with the corresponding terms from  
Fig.~\ref{diags4} in the singlet ($\sigma_{S, S}^{(2)}$ in Eq.~(\ref{sdecomp})),
non-singlet ($\sigma_{S, NS}$ and $\sigma_{A, NS}$ in Eqs.~(\ref{sdecomp}) 
and (\ref{adecomp})) or triangle ($\sigma_{S, Tr}^{(2)}$ and
$\sigma_{A, Tr}^{(2)}$ in Eqs.~(\ref{sdecomp}) and (\ref{adecomp}))
contributions.
The squared diagrams of type~$ii)$, which give a universal (i.e.~factorizable) 
correction to both the antisymmetric and symmetric cross sections, 
can be considered part of the non-singlet contributions.

It is not entirely clear how four-quark final states are 
actually treated in the different
experimental analyses, i.e.~the extent to which they are genuinely
measuring the inclusive cross sections.  
Often some vague statement
like ``a four-$b$ final state is more likely to be tagged than a two-$b$
one, but less than twice as likely'' is made.  To know what to calculate
one must understand the corrections that are applied for this difference
in tagging efficiency, which are not usually explicitly stated in the
papers.  In the absence of a unique experimental procedure and of a 
definitive statement from the experiments
on what they are measuring, we make this ambiguity explicit by
multiplying the $E$-term by an arbitrary weight factor
$W_E$~\footnote{Note that we use the same normalization as in
  Ref.~[\ref{ERT}] (see also Eq.~(\ref{esint})) in which the $E$-term
  already includes an identical-particle factor of $1/(2!)^2$ because
  there are two identical quarks and two identical antiquarks in the
  final state.  Thus, when we set $W_E=2$ we actually include an overall
  factor of $W_E/(2!)^2=1/2!$.}.  An
inclusive definition would correspond to $W_E=2$ (each $b$ quark
contributing once), while an exclusive definition (the cross section for
events containing at least one $b$ quark) would correspond to $W_E=1$.
Since the forward--backward asymmetry is defined to be the asymmetry of a
differential cross section, it is clear that we must use the same cross
section definition in the numerator and denominator, i.e.~that $W_E$
must be the same in the symmetric and antisymmetric cross sections.

We return to the r\^ole of the weight factor $W_E$ after discussing the general
form of the non-singlet contributions.

\subsection{Non-singlet contributions}
\label{sec:nonsing}

Here we consider all the other contributions that have not yet been
treated, namely all the diagrams in Fig.~\ref{diags1} except those in
Fig.~\ref{diags1}d, the diagrams in
Figs.~\ref{diags2}a,~\ref{diags2}b,~\ref{diags3},~\ref{diags4}a
and~\ref{diags5}a, as well as the $E$-term
defined above. All these terms are included in the non-singlet components
$\sigma_{S, NS}$ and $\sigma_{A, NS}$ of Eqs.~(\ref{sdecomp}) 
and (\ref{adecomp}). Actually, introducing the weight factor $W_E$ for the
$E$-term, we can define the following symmetric and antisymmetric cross sections
\beeq
\label{sxswe}
\sigma_{S, NS}(W_E) &=& \sigma_{S, NS}(W_E=0) + W_E \;\sigma_{S}^{(0)}
\int E_S \;, \\
\label{axswe}
\sigma_{A, NS}(W_E) &=& \sigma_{A, NS}(W_E=0) + W_E \;\sigma_{A}^{(0)}
\int E_A \;, 
\eeeq
where $\int E_S$ and $\int E_A$ denote the integral of the symmetric and
antisymmetric $E$-term, respectively. We recall that the `truly' inclusive cross
sections in Eq.~(\ref{saxs}) correspond to the definition with $W_E=2$,
i.e.~$\sigma_{S, NS}= \sigma_{S, NS}(W_E=2)$ and
$\sigma_{A, NS}= \sigma_{A, NS}(W_E=2)$.

The $\cO(\as^2)$-calculation of the cross sections in
Eqs.~(\ref{sxswe},~\ref{axswe}) and of the corresponding
forward--backward asymmetry in the case 
of a finite $b$-quark mass is extremely complicated, and we are not able to
perform it. It is thus convenient to separate the calculation
into a piece that is finite in the massless limit and a simpler piece
that is not. Then, the (although, cumbersome) finite piece can be more easily 
computed in the massless approximation, while the simpler non-finite piece
can be computed in the massive theory.

It is possible to show (Appendix~A) that the inclusive definition, with
$W_E=2$, results in an antisymmetric cross section $\sigma_A$
(or, analogously, $\sigma_{A, NS}$)  that is finite in the
massless limit, at least at $\mathcal{O}(\as^2)$.  However, in the same 
limit, the inclusive symmetric cross section is divergent at 
$\mathcal{O}(\as^2)$, even if we only consider its non-singlet component.
The corrections to (the non-singlet component of) 
the forward--backward asymmetry itself must therefore also be divergent
in the massless limit.

This final statement remains true for {\em any\/} value of $W_E>0$.  For
example,
with \mbox{$W_E=1$}, the non-singlet part of the symmetric cross section is 
finite (see Eq.~(\ref{sxswe1})), but the antisymmetric cross section contains 
log\-arithmically-enhanced terms.

The divergences in the non-singlet components correspond to
logarithmically-enhanced terms $\as^2\ln Q^2/m_b^2$ coming from the $E$-term
in the triple-collinear limit, i.e.~when three fermions of the four-quark
final state become simultaneously parallel (Appendix~B).
The integral of the symmetric $E$-term is calculated for finite values of the
quark mass in Appendix~B.  Neglecting corrections of $\cO(m_b/Q)$, the final 
result is
\begin{equation}
 \label{ES}
 \int E_S = C_F\left( C_F- \frac{C_A}{2} \right)\left(\frac{\as}{2\pi}\right)^2
 \left[2\left(\frac{13}4-\frac{\pi^2}2+2\zeta_3\right)\ln\frac{Q^2}{m_b^2}
 -8.1790\pm0.0013\right].
\end{equation}
As expected from the singular behaviour in the triple-collinear limit,
the analytic coefficient in front of $\ln Q^2/m_b^2$ is proportional
to the integral of the non-singlet Altarelli--Parisi probability
$P_{q {\bar q}}^{NS}(z,\as)$ (see, for instance, the first paper in
Ref.~[\ref{CFP}]):
\beeq
\int_0^1 \d z \, P_{q {\bar q}}^{NS}(z,\as) &=& 
\left( \frac{\as}{2 \pi} \right)^2 C_F \left( C_F - \frac{1}{2} C_A \right)
\left( \frac{13}{4} - \frac{\pi^2}{2} + 2 \zeta_3  \right) \\
\label{SLcoef}
&\simeq& \left( \frac{\as}{2 \pi} \right)^2 C_F \left( C_F - \frac{1}{2} C_A 
\right) \;0.7193 \;\;.
\eeeq
The constant term in the square bracket on the right-hand side of
Eq.~(\ref{ES}) is the result of our numerical calculation.

Having pointed out that the symmetric $E$-term is divergent in the massless
limit, it is very simple to show how the divergence appears in the 
inclusive symmetric cross section. According to the definition of the
non-singlet component of $\sigma_{S}$, the virtual diagrams that contribute to
$\sigma_{S, NS}$ are exactly those that contribute to the $e^+e^-$ total cross
section. As for the real diagrams, they only differ by the 
contributions coming from the $E$-term. In the total cross section, the
$E$-term enters with a multiplicity factor $W_E=1$, and its divergence is
cancelled by that of the virtual diagrams. In the inclusive $b$-quark cross
section, the multiplicity factor of the $E$-term is $W_E=2$ and, thus,
the cancellation of the divergence with the virtual terms is spoiled.

This argument also allows us to directly compute the $\cO(\as^2)$-correction to
Eq.~(\ref{sxswe}). Exploiting the fact that the massless QCD correction
to $\sigma_{S, NS}(W_E=1)$ is equal to the correction $R_{e^+e^-}$ to the total
cross section, we write
\beq
\label{sxswe1}
\sigma_{S, NS}(W_E=1) = \sigma_{S}^{(0)} \left[ R_{e^+e^-} + \cO(\as^3) \right]
\;,
\eeq 
and, more generally,
\beq
\label{sxswe2}
\sigma_{S, NS}(W_E) = \sigma_{S}^{(0)}
\left[ R_{e^+e^-} + (W_E -1) \int E_S  + \cO(\as^3) \right] \;\;.
\eeq
Then, we obtain an explicit expression for $\sigma_{S, NS}(W_E)$ by simply
introducing in Eq.~(\ref{sxswe2}) our result in Eq.~(\ref{ES}) for $\int E_S$
and the well-known result~[\ref{Rhad2}] for $R_{e^+e^-}$:
\beeq
\label{ree}
R_{e^+e^-} &=& 1 + \frac{3}{4} \,C_F \, \frac{\as(Q^2)}{\pi}  \\
&+& \left(\frac{\as(Q^2)}{2 \pi}\right)^2 C_F
\left\{ - \frac{3}{8} C_F + C_A \left( \frac{123}{8} - 11 \zeta_3 \right)
+ T_R N_f \left( 4 \zeta_3 - \frac{11}{2} \right) \right\} + \cO(\as^3) 
\;,\nonumber
\eeeq
where $T_R=1/2$ and $N_f$ is the number of light flavours (e.g.~$N_f=5$ at LEP).

In particular, for the inclusive symmetric cross section we obtain
\beq
\label{sxsinc}
\sigma_{S, NS} = \sigma_{S, NS}(W_E=2) = \sigma_{S}^{(0)}
\left[ R_{e^+e^-} +  \int E_S  + \cO(\as^3) \right] \;.
\eeq

The explicit $\cO(\as^2)$-calculation of the antisymmetric cross section
$\sigma_{A, NS}$ and of the forward--backward asymmetry is described in the
next Section.

Note that our result in Eq.~(\ref{sxsinc}) for the inclusive symmetric cross 
section disagrees with the corresponding result of Ravindran and van
Neerven~[\ref{RvN}]. Their expression for the correction to the
symmetric cross section
($f_T+f_L$ in their Eqs.~(31) and~(32)) is equal to the result in 
Eq.~(\ref{ree}) for the $\cO(\as^2)$-correction to $R_{e^+e^-}$.
The disagreement thus regards the additional logarithmically-enhanced term
$\int E_S$ included in our expression.
The multiplicity of $b$-quarks is not required to be finite
in massless QCD (even in the non-singlet sector), and thus we cannot find any
reason why this logarithmically-enhanced term can disappear from the 
inclusive symmetric cross section.

The results of Ref.~[\ref{RvN}] for $\sigma_{S,NS} =
\sigma_{S,NS}(W_E=2)$ are based on the calculation of the
single-particle inclusive distribution performed in Refs.~[\ref{RivNNP}].
Our result is consistent with those in Refs.~[\ref{RivNNP}]. In fact, we
have evaluated the integral over the longitudinal-momentum fraction $z$
of the non-singlet coefficient function
$C_{S,q}^{NS}(z,\as(Q^2),Q^2/\mu_F^2) = C_{T,q}^{NS} + C_{L,q}^{NS}$,
computed there. This integral is proportional to $\sigma_{S,NS} =
\sigma_{S,NS}(W_E=2)$ in massless QCD after factorization of
collinearly-divergent contributions at the factorization scale $\mu_F$.
We find that the integral explicitly depends on $\ln Q^2/\mu_F^2$, thus
proving that $\sigma_{S,NS}$ is not finite in massless QCD. The
coefficient of $\ln Q^2/\mu_F^2$ exactly agrees with the coefficient of
$\ln Q^2/m_b^2$ in our Eqs.~(\ref{ES}) and~(\ref{sxsinc}).

\section{Calculation of the non-singlet contribution at \boldmath$\cO(\as^2)$}
\label{sec:calcn}

As discussed in Sect.~\ref{sec:nonsing}, the NNLO corrections to the non-singlet
component of the forward--backward asymmetry, $A_{FB, NS}$, are divergent in the
massless limit. The divergent behaviour remains true also if we abandon the 
fully inclusive definition by introducing the arbitrary weight $W_E$ for
the $E$-term. Thus, $A_{FB, NS}$ cannot be computed at $\cO(\as^2)$ by using
the massless approximation.

Nonetheless, since both $\sigma_{A, NS}(W_E=2)$ and $\sigma_{S, NS}(W_E=1)$
are finite when $m_b \to 0$, we can use the dependence on
$W_E$ to construct an unphysical observable that is finite in the
massless limit:
\begin{equation}
  \label{AFBfinite}
  A_{FB}^{(2);\mbox{\footnotesize finite}} \equiv
  \frac{\sigma_{A, NS}(W_E=2)}{\sigma_{S, NS}(W_E=1)} \;.
\end{equation}
The physical result for $W_E=2$ is then given by
\begin{equation}
  \label{AFBfull}
  A_{FB, NS}^{(2)} = A_{FB}^{(2);\mbox{\footnotesize finite}}
  - \frac{\sigma_A^{(0)}}{\sigma_S^{(0)}} \int E_S 
  = A_{FB}^{(2);\mbox{\footnotesize finite}} - A_{FB}^{(0)} \int E_S \;,
\end{equation}
where $\int E_S$ is the integral of the symmetric $E$-term, given in 
Eq.~(\ref{ES}).

The massless calculation of $A_{FB}^{(2);\mbox{\footnotesize finite}}$
can be performed in a similar way to the NLO calculation of Sect.~\ref{sec:nlo}.
The total contribution can be written as
\begin{equation}
  \label{ratio2}
  A_{FB}^{(2);\mbox{\footnotesize finite}} = \frac
  {\sigma_A^{(0)} + \sigma_A^{(1)}
    + \sigma_A^{(2);\mbox{\footnotesize two-loop}} 
    + \sigma_A^{(2);\mbox{\footnotesize one-loop}} 
    + \sigma_A^{(2);\mbox{\footnotesize tree}}(W_E=2)}
  {\sigma_S^{(0)} + \sigma_S^{(1)}
    + \sigma_S^{(2);\mbox{\footnotesize two-loop}} 
    + \sigma_S^{(2);\mbox{\footnotesize one-loop}} 
    + \sigma_S^{(2);\mbox{\footnotesize tree}}(W_E=1)} \;,
\end{equation}
where $\sigma_A^{(1)}$ and $\sigma_S^{(1)}$ are the complete contributions to
the antisymmetric and symmetric cross sections at $\cO(\as)$. The non-singlet
$\cO(\as^2)$-contributions from the two-parton, three-parton and four-parton
final states are denoted by $\sigma^{(2);\mbox{\footnotesize two-loop}}, 
\sigma^{(2);\mbox{\footnotesize one-loop}}$ and 
$\sigma^{(2);\mbox{\footnotesize tree}}$ respectively.
Of course, the dependence on $W_E$
enters only through the four-parton terms 
$\sigma_A^{(2);\mbox{\footnotesize tree}}(W_E=2)$ and
$\sigma_S^{(2);\mbox{\footnotesize tree}}(W_E=1)$.

If we continue to use a regularization scheme that preserves the
helicity conservation of massless QCD, like dimensional regularization,
the two-loop corrections are again proportional to the tree-level
results,
\begin{equation}
  \frac{\sigma_A^{(2);\mbox{\footnotesize two-loop}}}{\sigma_A^{(0)}}
  =\frac{\sigma_S^{(2);\mbox{\footnotesize two-loop}}}{\sigma_S^{(0)}} \;,
\end{equation}
so that if we expand the ratio in Eq.~(\ref{ratio2}) up to 
$\mathcal{O}(\as^2)$, the two-loop corrections cancel, and we obtain
\begin{eqnarray}
  A_{FB}^{(2);\mbox{\footnotesize finite}} 
  &=& \!\! \frac{\sigma_A^{(0)}}{\sigma_S^{(0)}}
  \Biggl[1+
  \Biggl(1-
  \frac{\sigma_S^{(1)}}{\sigma_S^{(0)}}
  \Biggr)\Biggl(
   \frac{\sigma_A^{(1)}}{\sigma_A^{(0)}} -
   \frac{\sigma_S^{(1)}}{\sigma_S^{(0)}}
  \Biggr)
\nonumber\\\label{afb2exp} && \hspace{-1.5em}
  + \frac{\sigma_A^{(2);\mbox{\footnotesize one-loop}}}{\sigma_A^{(0)}} -
  \frac{\sigma_S^{(2);\mbox{\footnotesize one-loop}}}{\sigma_S^{(0)}}
  + \frac{\sigma_A^{(2);\mbox{\footnotesize tree}}(W_E=2)}{\sigma_A^{(0)}} -
  \frac{\sigma_S^{(2);\mbox{\footnotesize tree}}(W_E=1)}{\sigma_S^{(0)}}
  \Biggr] \,.
\end{eqnarray}
The first line can be calculated analytically (see Sect.~\ref{sec:nlo}), 
but the second line is
too complicated to be able to, so must be done numerically.  Since the
two-loop terms have cancelled, this has the structure of a NLO three-jet
calculation, as first noticed by Altarelli and Lampe~[\ref{AL}]. Thus
the calculation can be performed
using known techniques (we use the dipole-formalism version of the 
subtraction method~[\ref{CS}]).
One simply has to replace the full matrix element
squared by the appropriate contractions of the hadronic tensor, as in
Eqs.~(\ref{tensorS}) and~(\ref{tensorA}).
We have obtained simplified analytical expressions for these
contractions by using the matrix elements originally computed by 
the Leiden group~[\ref{ZvN}]. We have also checked that these expressions
numerically agree with the code of Ref.~[\ref{GG}].

\subsection{Numerical results}
\label{sec:numres}

We are finally ready to present our numerical results.  We start with
the unphysical, but finite, quantity defined in Eq.~(\ref{AFBfinite}),
and separate out the different colour factors, as in
Refs.~[\ref{AL},~\ref{RvN}]:
\begin{eqnarray}
  \label{AFBbCNTdef}
  A_{FB}^{(2);\mbox{\footnotesize finite};b} &=&
  A_{FB}^{(0)}\left[
    1-\frac{\as}{2\pi}\left(1-\frac{\as}{2\pi}\frac32C_F\right)
    \left(\frac32C_F\right)
    \phantom{\left(\frac{\as}{2\pi}\right)^2}
    \phantom{\left(\frac{\as}{2\pi}\right)^2}\right.
\nonumber\\&&
  \phantom{\left(\frac{\as}{2\pi}\right)^2}
  \phantom{\left(\frac{\as}{2\pi}\right)^2}
  \left.
    +\left(\frac{\as}{2\pi}\right)^2C_F\left(
      CC_F+NN_C+TT_RN_f\right)\right] \;,
\end{eqnarray}
with $\as \equiv \as(Q^2)$.
Our numerical results are shown in Table~\ref{table1}, in comparison
with the previous calculations.
\begin{table}[b]
 \begin{center}
  \begin{tabular}{|c|c|c|c|}
    \hline
    $b$-quark axis & $C$ & $N$ & $T$ \\
    \hline
    AL [\ref{AL}] & $4.4\pm0.5$ & $-10.3\pm0.3$ & $5.68\pm0.04$ \\
    RvN [\ref{RvN}] & $\frac38 = 0.375$ & $-\frac{123}8 = -15.375$ &
      $\frac{11}2 = 5.5$ \\
    \hline
    \hline
    Our Calculation & $0.3765\pm0.0038$ & $-15.3769\pm0.0034$ &
      $5.5002\pm0.0008$ \\
    \hline
  \end{tabular}
 \end{center}
 \caption{Results for the coefficients of the $\cO(\as^2)$ correction to
    the finite part of the forward--backward asymmetry with the $b$-quark
    axis definition, Eqs.~(\ref{AFBfull},~\ref{AFBbCNTdef}).}
 \label{table1}
\end{table}
It is clear that we disagree badly with the results of Altarelli and
Lampe~[\ref{AL}], but are in excellent agreement with Ravindran and van
Neerven~[\ref{RvN}], who give the coefficients analytically.
However, we should recall that this must have subtracted from it the
logarithmically-enhanced term of Eqs.~(\ref{AFBfull},~\ref{ES}), which is not
present in the result of Ref.~[\ref{RvN}]. In fact, in Sect.~\ref{sec:nonsing}
we have already pointed out that their expression
for the correction to the symmetric cross section does not agree with ours,
but, rather, it is actually equal to our $\sigma_{S,NS}^{(2)}(W_E=1)$.
So, the fact that their result for the complete $A_{FB}^{(2)}$ agrees
with our $A_{FB}^{(2);\mbox{\footnotesize finite}}$ means that we
confirm their result~[\ref{RivN},~\ref{RvN}]
for the inclusive antisymmetric cross section 
$\sigma_A^{(2)}=\sigma_A^{(2)}(W_E=2)$ 
($f_A$ in Eq.~(33) of Ref.~[\ref{RvN}]).

The disagreement with the result of Ref.~[\ref{AL}] may be related to
the poor numerical convergence of their calculational method (i.e.~the
effect of large numerical cancellations).

Using our numerical program it is straightforward to calculate the
forward--backward asymmetry with any other axis definition (or cuts, for
example on the value of the thrust).  With the thrust axis definition,
we obtain
\begin{eqnarray}
  \label{AFBTCNTdef}
  A_{FB}^{(2);\mbox{\footnotesize finite};T} &=&
  A_{FB}^{(0)}\left[
    1-\frac{\as}{2\pi}\left(1-\frac{\as}{2\pi}\frac32C_F\right)
    \left(1.34 C_F \right)
    \phantom{\left(\frac{\as}{2\pi}\right)^2}
    \phantom{\left(\frac{\as}{2\pi}\right)^2}\right.
\nonumber\\&&
  \phantom{\left(\frac{\as}{2\pi}\right)^2}
  \phantom{\left(\frac{\as}{2\pi}\right)^2}
  \left.
    +\left(\frac{\as}{2\pi}\right)^2C_F\left(
      CC_F+NN_C+TT_RN_f\right)\right],
\end{eqnarray}
with $\as \equiv \as(Q^2)$ and the coefficients given in Table~\ref{table2}.
\begin{table}[t]
 \begin{center}
  \begin{tabular}{|c|c|c|c|}
    \hline
    thrust axis & $C$ & $N$ & $T$ \\
    \hline
    \hline
    Our Calculation & $-3.7212\pm0.0065$ & $-9.6011\pm0.0049$ &
      $4.4144\pm0.0006$ \\
    \hline
  \end{tabular}
 \end{center}
 \caption{Results for the coefficients of the $\cO(\as^2)$ correction to
    the finite part of the forward--backward asymmetry with the thrust
    axis definition, Eqs.~(\ref{AFBfull},~\ref{AFBTCNTdef}).}
 \label{table2}
\end{table}
The logarithmically-enhanced piece that has to be added to this is identical
to that in the $b$-quark axis definition, namely
Eqs.~(\ref{AFBfull},~\ref{ES}).  It is worth noting that the difference
between the two definitions is the same size and in the same direction
as at $\mathcal{O}(\as)$, leading to an overall difference of 0.8\% for 
$\as \sim 0.12$.

Since $A_{FB}^{(2);\mbox{\footnotesize finite};T}$ is defined by the ratio
in Eq.~(\ref{AFBfinite}), using the expression in Eq.~(\ref{sxswe1}) for 
$\sigma_{S, NS}(W_E=1)$, we can translate our result in Eq.~(\ref{AFBTCNTdef})
into an equivalent result for the antisymmetric cross section defined with
respect to the thrust axis. We have:
\beeq
\label{sigANST}
\sigma_{A, NS}^{T} &=& \sigma_A^{(0)}\Biggl\{ 1 + 0.034 \;\as(Q^2)  
+ \left(\frac{\as(Q^2)}{2\pi}\right)^2 C_F
\left[ \left( -\frac{3}{8} +C \right) C_F  \right. \nonumber \\
\label{satnnlo}
&& \left.  + \left( \frac{123}{8} - 11 \zeta_3 + N \right) C_A
      + \left( 4 \zeta_3 - \frac{11}{2} + T \right) T_R N_f
\right] + \cO(\as^3) \Biggr\} \,.
\eeeq
with the coefficients $C, N$ and $T$ given in Table~\ref{table2}~\footnote{In 
the analogous expression for $\sigma_{A,NS}^{b}$, the coefficient of $\as(Q^2)$
vanishes and $C, N$ and $T$ are those given in Table~\ref{table1}, which
exactly cancel the rational numbers in Eq.~(\ref{sigANST}), leaving only
$3\beta_0\zeta_3$, with $\beta_0=\frac{11}3C_A-\frac43T_RN_f$, as pointed
out in Ref.~[\ref{RivN}].}.

We finally recall that we include an arbitrary factor $W_E$ in front of
the four-$b$ contribution to account for the way in which it is treated
in the experimental analyses.  For a fully inclusive definition, in
which each $b$ quark contributes once, $W_E$ should be set equal to 2,
while for an exclusive definition, $W_E$ should be set equal to 1.  Our
final result for the non-singlet component of the forward--backward
asymmetry, is then:
\begin{equation}
  \label{AFBfullwe}
  \hspace*{-1em}
  A_{FB, NS}^{(2)}(W_E) \equiv 
  \frac{\sigma_{A, NS}(W_E)}{\sigma_{S, NS}(W_E)} =
   A_{FB}^{(2);\mbox{\footnotesize finite}} - A_{FB}^{(0)}
   \left[ (1-\smfrac12W_E) \left( 2\int E_A - \int E_S \right)
   + \smfrac12W_E \int E_S \right] \;,
  \hspace*{-1em}
\end{equation}
where $A_{FB}^{(2);\mbox{\footnotesize finite}}$ is given in
Eqs.~(\ref{AFBbCNTdef},~\ref{AFBTCNTdef}) and Tables~\ref{table1}
and~\ref{table2}, $\int E_S$ is given in Eq.~(\ref{ES}) and (see
Appendix~B)
\beeq
\label{EASintb}
2\int E_A - \int E_S &=& \left(\frac{\as}{2\pi}\right)^2
C_F(C_F-\smfrac12C_A)\Bigl(0.3620\pm0.0007\Bigr),\qquad\mbox{quark axis}, \\
\label{EASintT}
2\int E_A - \int E_S &=& \left(\frac{\as}{2\pi}\right)^2
C_F(C_F-\smfrac12C_A)\Bigl(0.1144\pm0.0009\Bigr),\qquad\mbox{thrust axis}.
\eeeq
Note that the combinations of $E$-term contributions in Eqs.~(\ref{EASintb}) 
and (\ref{EASintT}) are finite in the massless limit (see the discussion in 
Appendix~B).

Putting all these numbers together, and setting $N_f=5$, we write the
forward--backward asymmetry according to the two definitions as:
\begin{eqnarray}
  A_{FB, NS}^{(2);b}(W_E) &=& A_{FB}^{(0)} \left[
  1-0.318\as-0.973\as^2+W_E\as^2\left(0.00405\ln\frac{Q^2}{m_b^2}-0.0240\right)
  \right], \\
  A_{FB, NS}^{(2);T}(W_E) &=& A_{FB}^{(0)} \left[
  1-0.284\as-0.676\as^2+W_E\as^2\left(0.00405\ln\frac{Q^2}{m_b^2}-0.0233\right)
  \right].\phantom{(99)}
\end{eqnarray}

\section{Conclusion}
\label{sec:concl}

We have calculated the second-order corrections to the non-singlet
component of the forward--backward asymmetry in $e^+e^-$ annihilation.
We have retained all terms that do not vanish in the small-mass limit
(constants and logarithmically-enhanced terms).  Our result is also
valid for the left--right forward--backward asymmetry.

Using the quark axis definition we do not agree with any existing
calculation.  Separating the asymmetry into its symmetric and
antisymmetric parts, we find that we agree with Ravindran and van
Neerven~[\ref{RvN}] for the antisymmetric part, which is finite in
massless QCD\@.  For the symmetric part we disagree by a term that is
divergent in massless QCD, so is logarithmically-enhanced in the full
theory.

We have obtained results for the first time with the thrust axis
definition, which is actually what is used in the experimental
measurements.  After including the second-order contributions, the
difference between the two axis definitions is twice as large as at
first order, amounting to 0.8\%.

We summarize the total QCD correction according to the various available
calculations in Table~\ref{tab:concl}.
\begin{table}[b]
 \begin{center}
  \begin{tabular}{|c|c|c|c|c|}
    \hline
    & AL [\ref{AL}] & RvN [\ref{RvN}] & Our Calculation & Our Calculation \\
    & quark axis & quark axis & quark axis & thrust axis \\
    \hline
    Correction, $A_{FB}^{(2)}/A_{FB}^{(0)}$ &
    0.962 &
    0.952 &
    0.952 &
    0.956 \\
    \hline
  \end{tabular}
 \end{center}
 \caption{Total QCD correction to the forward--backward asymmetry in the
    small-mass limit, with $\as=0.12$.  In each case, the thrust axis
    definition is used for the $\cO(\as)$ correction and the definition
    shown is used for the $\cO(\as^2)$ correction, as discussed in the
    text.}
 \label{tab:concl}
\end{table}
We continue to neglect all terms that vanish in the massless limit, and
discuss the effect of mass corrections below.  Since in the existing
experimental analyses (see for example Ref.~[\ref{abbaneo}]), the known
$\cO(\as)$ correction for the thrust axis definition was included,
together with the Altarelli and Lampe quark axis value for the
$\cO(\as^2)$ corrections, we do the same in Table~\ref{tab:concl}.

We find that the difference between the Ravindran and van Neerven
calculation and ours is numerically irrelevant, being smaller than
$10^{-4}$ for $b$ quarks and $\sim2.5\!\times\!10^{-4}$ for $c$ quarks.
Therefore at the numerical precision required by current or any foreseen
experiments, we agree with their result~-- the difference is only one of
principle.  The difference between the Altarelli and Lampe calculation
and ours for the quark axis definition is more significant though, at
around 1\%.  However, the error in their calculation and the effect of
using the thrust axis definition partially cancel, and the total
difference is around 0.6\%.

Before quantifying the impact of these differences, we mention the
important fact, discussed in Ref.~[\ref{abbaneo}], that the experimental
procedures introduce a bias towards more two-jet-like events.  This
actually decreases the size of the QCD corrections considerably, so our
numbers should be considered as upper bounds.  In fact at present the
effect of this bias is typically taken into account using Monte Carlo
event generators.  Using our numerical calculation, it is
straightforward to apply any infrared-safe cuts, for example on the
thrust of the event (this effect was first considered at $\cO(\as)$ in
Ref.~[\ref{DLZ}]).  This could be used to reduce the reliance on the
generators, or at the least to calibrate their reliability.

To quantify the impact of the differences shown in
Table~\ref{tab:concl}, we recall a few figures
from the latest global electroweak fit~[\ref{electroweak}].  The total
error on the LEP average forward--backward asymmetry of $b$-quarks
$A_{FB}$ is 2.1\%.  The second-order QCD corrections are used to convert
the measured value into a measurement of the tree-level asymmetry,
$A_{FB}^{(0)}$, at present using the Altarelli and Lampe result.  This
is then used as input into the fit for the effective weak mixing angle,
$\sin^2\theta_{ef\!f}$, and eventually into the global fit to all
electroweak data.  Following through this process, our smaller value of
the correction in Table~\ref{tab:concl} results in a larger value of
$A_{FB}^{(0)}$ and hence a smaller value of $\sin^2\theta_{ef\!f}$, by
about a third of its experimental error.

This has a direct bearing on the fitted value of the Higgs mass, (see
Fig.~9 of Ref.~[\ref{electroweak}]).  To find the effect of using our
corrections would require a complete refitting of all the electroweak
data.  However, we can get a rough idea simply by fitting the data in
Fig.~9 of Ref.~[\ref{electroweak}] alone.  We find a roughly linear
relation: for each {\em per mille\/} that the corrected value of the
quark asymmetries is {\em increased}, we obtain a {\em per cent} {\em
  decrease\/} in the central value of the Higgs mass (and its upper
bound).  Therefore with our 0.6\% difference, we expect a reduction of
about 5~GeV in the central value.  While this is certainly not
statistically significant, given the importance that some people attach
to this value, it is not irrelevant either.

In trying to estimate the remaining uncertainties in the
forward--backward asymmetry, we recall the ingredients still missing from
our analysis.  We should bear in mind that while the 2\% precision of
current experiments is close to their final limit, a future linear
collider could be capable of experimental errors on the left--right
forward--backward asymmetry of order 0.1\%~[\ref{NLC}].

Within small-mass perturbation theory, the first terms that we neglect
are $\cO(\as^3)$.  To estimate their size, we assume that the
coefficient grows as much in going from $\cO(\as^2)$ to $\cO(\as^3)$ as
it did from $\cO(\as)$ to $\cO(\as^2)$, and get 0.3\%.  The more
conventional method, varying the renormalization scale from $Q/2$ to
$2Q$ results in a similar estimate of 0.2\%.  A variation in the input
value of $\as$ of $\pm0.004$ gives only 0.1\%.

Within the $\cO(\as^2)$ calculation, we neglected the effect of triangle
diagrams.  For the quark axis definition, these were calculated in
Ref.~[\ref{AL}], and amount to about 0.1\%.  We have no reason to
suppose they would be larger for the thrust axis definition, and in any
case it would not be difficult to calculate them.

We have also neglected linear mass corrections of the type $m_b/Q$,
which are absent at tree level, but arise at higher orders.  The full
mass correction at $\cO(\as)$ is well known, and is reasonably well
approximated by its leading term,
$4C_F\as/\pi\,m_b/Q$.  Since we do not have any higher order corrections to
this linear mass term, its renormalization group dependence is not under
control, so to estimate the effect of higher order corrections, we vary
$m_b$ from its running value in the $\overline{\mathrm{MS}}$ scheme
($\sim3$~GeV) to its pole
value ($\sim5$~GeV), resulting in a 0.4\% variation in $A_{FB}$.

Finally, at higher orders it is quite possible that the leading mass
term could become logarithmically enhanced, for instance, as $\sim\as^2
m_b/Q\ln^n(Q^2/m_b^2)$ at the second order.  Terms like this certainly
arise with $n\!=\!1$ simply from the renormalization group effects just
mentioned, but the question is whether additional terms can arise from
other dynamic effects.  A possible additional source of
single-logarithmic enhancement is collinear emission, as in the case of
the $E$-term contributions discussed earlier.  Owing to the
inclusiveness of the forward--backward asymmetry with respect to soft
emission, we think that higher powers of logs are unlikely to be present
in the non-singlet component at $\cO(\as^2)$.  Although this point
deserves further investigation, assuming $n \leq 1$ we estimate a
resulting uncertainty of 0.5\%.

We have not made any attempt to estimate the uncertainty due to
non-perturbative corrections.  In Ref.~[\ref{abbaneo}], this is done
using Monte Carlo event generators.  They find a correction of 0.25\%
and conservatively assign the whole of this as a systematic error.

To summarize, there are several sources of uncertainty that all
contribute at the few per mille level.  While this is certainly
sufficient for the current precision of the data, matching the precision
of a future linear collider measurement could be extremely difficult.
It is likely that this could only be done by making even more stringent
two-jet cuts in order to work in a region in which the corrections and
their uncertainties are smaller.

\vspace{-2ex}
\subsubsection*{Acknowledgements}
\vspace{-1ex}
We are grateful to Guido Altarelli, Klaus M\"onig and especially Willy
van Neerven for discussions of the forward--backward asymmetry and
comments on the manuscript.  MHS gratefully acknowledges the hospitality
of the Theoretical Physics Department of Lund University during the
completion of this work.

\appendix
\renewcommand{\theequation}{A.\arabic{equation}}
\setcounter{equation}{0}
\subsection*{Appendix A: The antisymmetric cross section in massless QCD}

In this Appendix we show that, up to $\cO(\as^2)$, the perturbative-QCD 
corrections to the heavy-quark antisymmetric cross section $\sigma_A$ are 
finite in the limit of vanishing quark masses.

We are interested in the analogue of the cross section in Eq.~(\ref{dcs})
for the inclusive process $e^+e^- \to a + X$ where $a=q_f, {\bar q}_f, g$
denotes a generic massless QCD parton. We thus define the
antisymmetric\footnote{Exactly analogous definitions hold for the
symmetric cross section $\d\sigma_S^{a}/\d x$ and its $N$-moments
$\sigma_{S, N}^{a}$.} cross section $\d\sigma_A^{a}/\d x$ as follows
\beq
\label{axsx}
\frac{\d\sigma_A^a}{\d x} =\int_{0}^{1} \d\!\cos\theta \,
\frac{\d\sigma(e^+e^- \to a + X)}{\d x \;\d\!\cos\theta}
- \int_{-1}^{0} \d\!\cos\theta \,
\frac{\d\sigma(e^+e^- \to a + X)}{\d x \;\d\!\cos\theta} \;\;.
\eeq
It is also convenient to introduce the $N$-moments $\sigma_{A, N}^{a}$
defined by
\beq
\label{xsn}
\sigma_{A, N}^{a} = \int_0^1 \d x \; x^{N-1} \;\frac{\d\sigma_A^a}{\d x} \;\;,
\eeq
and likewise for any other function of the energy fraction $x$. Note
that the massless limit of the $b$-quark antisymmetric cross section in 
Eq.~(\ref{saxs}) coincides with the $N=1$ moment of $\d\sigma_{A}^{q_f}/\d x$,
i.e.~$\sigma_A = \sigma_{A, N=1}^{q_f}$.

Since we are working in massless QCD, the antisymmetric cross section
$\d\sigma_A^{a}/\d x$ is not finite in perturbation theory and, more precisely,
it is collinear divergent. Nonetheless, because of the
factorization theorem of mass singularities, once the divergences have been
regularized (by using, for instance, dimensional regularization) they
can be factorized. The $N$-moments can be written as
\beq
\label{sigman}
\sigma_{A, N}^{a} = \sum_{b=q_f, \,{\bar q}_f, g} {\hat \sigma}_{A, N}^{b}
\;\Gamma_{ba, N} \;\;,
\eeq
where ${\hat \sigma}_{A, N}^{b}$ is a finite contribution to the
cross section and
the factor $\Gamma_{ab, N}$ contains all the collinear singularities
(see e.g.~Ref.~[\ref{CFP}]).
This factor depends on the factorization (or regularization)
scale $\mu$ and the factorization scheme, but it is universal 
(process independent).
Moreover, it fulfils the Altarelli--Parisi evolution equations 
\beq
\label{apeq}
\frac{\partial}{\partial \ln \mu^2} \;\Gamma_{ab, N} = \sum_c 
P_{ac, N}(\as(\mu^2)) \; \Gamma_{cb, N} \;\;,
\eeq
with the initial condition $\Gamma_{ab, N}(\mu^2=0) = \delta_{ab}$ and
where $P_{ac, N}(\as)$ are the $N$-moments of the Altarelli--Parisi
probabilities, whose power series expansion in $\as$ can be computed at any 
perturbative order.

Note that the antisymmetric cross section
$\sigma_{A, N}^{a}$ is an odd quantity under charge conjugation.
Thus we have $\sigma_{A, N}^{a}= - \sigma_{A, N}^{{\bar a}}$ and, in particular,
$\sigma_{A, N}^{q_f}= - \sigma_{A, N}^{{\bar q_f}}$ and 
$\sigma_{A, N}^{g}= 0$. Analogous relations are valid for 
${\hat \sigma}_{A, N}^{a}$.

We can now consider in detail
the massless limit of the $b$-quark antisymmetric cross section
$\sigma_A$, that
is, the first moment $\sigma_{A, N=1}^{q_f}$. 
Since  ${\hat \sigma}_{A, N}^{a}$ is $C$-odd, Eq.~(\ref{sigman}) gives
\beq
\label{sigmanf}
\sigma_{A, N=1}^{q_f} = \sum_{f^\prime} {\hat \sigma}_{A, N=1}^{q_{f^\prime}}
\;\Gamma_{f^\prime f, N=1} \;\;,
\eeq
where
\beq
\Gamma_{f^\prime f, N=1} \equiv  \Gamma_{q_{f^\prime}q_f, N=1} - 
\Gamma_{{\bar q}_{f^\prime}q_f, N=1} \;\;.
\eeq
Using Eq.~(\ref{apeq}) and the property 
$P_{ac}(\as) = P_{{\bar a}{\bar c}}(\as)$, which follows from 
the charge-conjugation invariance of QCD,
we obtain the following evolution equation for the singular collinear factor
on the right-hand side of Eq.~(\ref{sigmanf})
\beq
\label{eveq}
\frac{\partial}{\partial \ln \mu^2} \,\Gamma_{f' f, N=1}
= \sum_{f''}
\left[ P_{q_{f'}q_{f''}, N=1}(\as(\mu^2)) - 
P_{{\bar q}_{f'}q_{f''}, N=1}(\as(\mu^2))
\right] \;\Gamma_{f'' f, N=1}  \;\;,
\eeq
Note that the combination of first moments of the Altarelli--Parisi
probabilities on the right-hand side of Eq.~(\ref{eveq}) vanishes up to
$\cO(\as^2)$:
\beq
\label{apcom}
P_{q_{f'}q_{f''}, N=1}(\as(\mu^2)) - 
P_{{\bar q}_{f'}q_{f''}, N=1}(\as(\mu^2))
= \cO(\as^3) \;\;.
\eeq
This result follows from fermion-number conservation and
it can be explicitly checked by using
the known LO and NLO expressions~[\ref{CFP}] of the Altarelli--Parisi
probabilities. Equation (\ref{apcom}) implies that
$\Gamma_{f^\prime f, N=1} = \delta_{f'f} + \cO(\as^3)$
and, thus, the massless-quark antisymmetric cross section
$\sigma_{A, N=1}^{q_f}$
is free from collinear singularities up to NNLO accuracy:
\beq
\label{sigmanf2}
\sigma_{A, N=1}^{q_f} = {\hat \sigma}_{A, N=1}^{q_{f}} + \cO(\as^3) \;\;.
\eeq

To conclude our argument on the finiteness of $\sigma_{A, N=1}^{q_f}$, we have
to discuss the effect of soft singularities. The QCD factorization theorem
guarantees that the short-distance cross section 
${\hat \sigma}_{A, N}^{q_{f}}$ is finite for any value of the moment
index $N > 1$. The limit $N \to 1$ of Eqs.~(\ref{xsn}) and (\ref{sigman})
has to be dealt with with care because it is
sensitive to possible soft singularities of the type 
$\d\sigma_A^{q_f}/\d x \sim 1/x$ in the inclusive quark spectrum.
These singularities can arise when $q_f$ is produced by the fragmentation of a
soft gluon. At $\cO(\as)$ there are no such fragmentation subprocesses. At
$\cO(\as^2)$, there is only the subprocess $g \to q_f {\bar q}_f$. 
In this $\cO(\as^2)$-subprocess, however, the pair $q_f {\bar q}_f$ is 
produced in a definite state of positive charge conjugation and, thus,
it gives a
vanishing contribution to the $C$-odd cross section $\sigma_A^{q_f}$. It 
follows that up to $\cO(\as^2)$ the limit $N \to 1$ can be safely 
performed and the right-hand side of Eq.~(\ref{sigmanf2}) is finite.

The antisymmetric cross section $\sigma_A$ is the integral of a single-particle
(parton) inclusive distribution and thus, the finiteness of 
Eq.~(\ref{sigmanf2}) may appear surprising. However, this result is not
accidental. The collinear safety of $\sigma_{A, N=1}^{q_f}$ follows from
fermion-number conservation, i.e.~Eq.~(\ref{apcom}), and this is exactly the
same equation that, up to $\cO(\as^2)$, guarantees the finiteness of the
Adler, Gross--Llewellyn-Smith and unpolarized-Bjorken sums in
Deep-Inelastic-Scattering.

\renewcommand{\theequation}{B.\arabic{equation}}
\setcounter{equation}{0}
\subsection*{Appendix B: Integrating the \boldmath$E$-terms}

In this Appendix we give some details of the integration of the
$E$-terms appearing in, for example, Eq.~(\ref{AFBfullwe}).

We begin with the symmetric term $\int E_S$, which is equal to the total
contribution to $R_{e^+e^-}$ from four-$b$ final states.  We label the
quark (antiquark) momenta as $p_{1,2}\,(p_{3,4})$ and retain all mass
terms.  The integral is then analogous to Eq.~(B.2) of Ref.~[\ref{ERT}], but
with the massless phase space replaced by that with four equal
final-state masses:
\begin{eqnarray}
\label{esint}
  \int E_S &=& \frac1{(2!)^2} C_F \left(\frac{\as}{2\pi}\right)^2 \frac1{Q^2}
  \int d\Phi_4(Q^2;m_b^2,m_b^2,m_b^2,m_b^2)
\nonumber\\&&
  \Biggl[ E_S(p_1,p_2,p_3,p_4) + (p_1\leftrightarrow p_2)
    + (p_3\leftrightarrow p_4)
    + (p_1\leftrightarrow p_2,p_3\leftrightarrow p_4)
  \Biggr],
\end{eqnarray}
and with the $E$-term itself in Eq.~(B.7) of Ref.~[\ref{ERT}] replaced
by:
\\\vbox{\begin{eqnarray}
  E_S &=& (C_F-C_A/2)\Biggl\{\Biggl[\biggl(
    (s_{12}s_{23}s_{34} - s_{12}s_{24}s_{34} + s_{12}s_{14}s_{34} +
    s_{12}s_{13}s_{34} + s_{13}s_{24}^2
\nonumber\\&&\qquad
    - s_{14}s_{23}s_{24} + s_{13}s_{23}s_{24} + s_{13}s_{14}s_{24} +
    s_{13}^2s_{24} - s_{14}s_{23}^2 - s_{14}^2s_{23} -
    s_{13}s_{14}s_{23})
\nonumber\\&&
    - 2m_b^2  ( 2s_{12}s_{13} + 3s_{12}s_{14} + s_{12}s_{23} - s_{12}s_{24}
    - s_{12}s_{34} + 2s_{13}s_{14} + s_{13}s_{24}
\nonumber\\&&\qquad
    + s_{13}^2 + s_{14}s_{23} + s_{14}s_{24} + s_{14}s_{34} -
    s_{23}s_{24} - s_{23}s_{34} - s_{23}^2 - 3s_{24}s_{34} - s_{34}^2 )
\nonumber\\&&
    + 4m_b^4  ( s_{12} + 2s_{13} + 5s_{14} + s_{23} - 2s_{24} - 3s_{34} )
\nonumber\\&&
    - 16m_b^6\biggr)/(s_{13}s_{23}s_{123}s_{134})
    -\biggl(
    s_{12}( s_{12}s_{34} - s_{23}s_{24} - s_{13}s_{24} - s_{14}s_{23}
    - s_{14}s_{13} )
\nonumber\\&&
    - 2m_b^2  ( 4s_{12}s_{34} + 2s_{12}^2 - 2s_{13}s_{14} - 2s_{13}s_{23}
    - 2s_{13}s_{24} - s_{13}^2 - 2s_{14}s_{23} - 2s_{23}s_{24} -
    s_{23}^2 )
\nonumber\\&&
    - 4m_b^4  (  - 6s_{12} + 2s_{13} + s_{14} + 2s_{23} + s_{24} - 3s_{34} )
\nonumber\\&&
    - 24m_b^6\biggr)/(s_{13}s_{23}s_{123}^2)
    -\biggl(
    (s_{14}+s_{13})(s_{24}+s_{23})s_{34}
\nonumber\\&&
    - m_b^2  ( 2s_{12}s_{13} + 2s_{12}s_{23} + s_{12}s_{34} + s_{12}^2 +
    4s_{13}s_{23} + s_{13}s_{24}
\nonumber\\&&\qquad
    + s_{13}s_{34} + s_{13}^2 + s_{14}s_{23} - s_{14}s_{34} +
    s_{23}s_{34} + s_{23}^2 - s_{24}s_{34} )
\nonumber\\&&
    - 2m_b^4  (  - 2s_{12} - s_{13} + 3s_{14} - s_{23} + 3s_{24} + 2s_{34} )
\nonumber\\&&
    + 24m_b^6\biggr)/(s_{13}s_{23}s_{134}s_{234})\Biggr]
\nonumber\\&&
    + \Biggl[(p_1\leftrightarrow p_3,p_2\leftrightarrow p_4)\Biggr]\Biggr\},
\end{eqnarray}}\\
where $s_{ij}=(p_i+p_j)^2$ and $s_{ijk}=(p_i+p_j+p_k)^2-m_b^2$.  Note
that by setting $m_b=0$ we trivially recover the result of
Ref.~[\ref{ERT}].

\begin{figure}
  \centerline{\resizebox{0.8\textwidth}{!}{\includegraphics{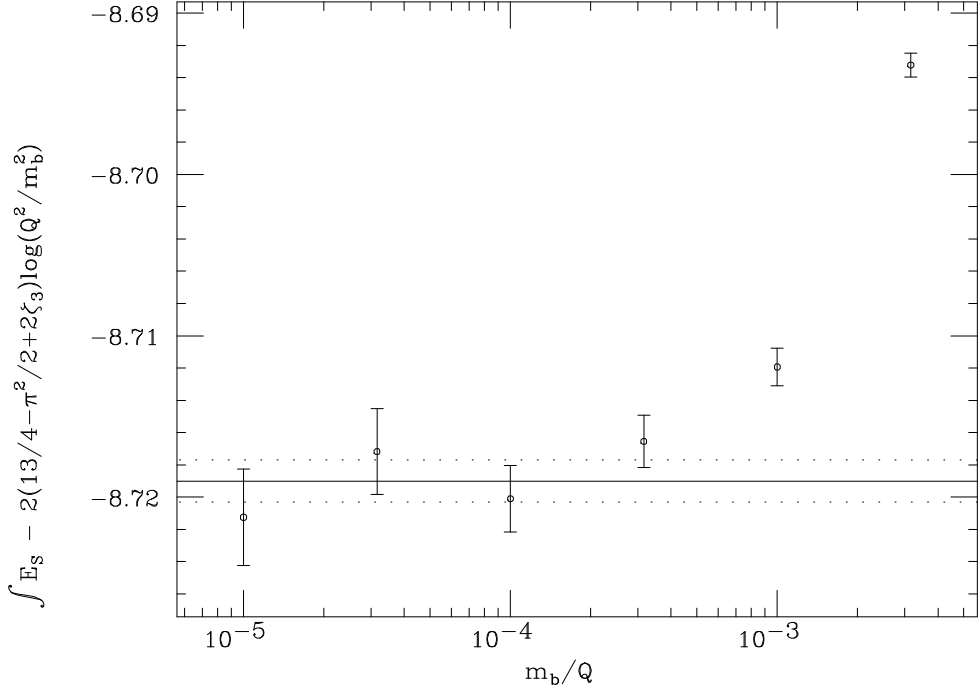}}}
  \caption{The term $c$ in Eq.~(\ref{Eint}) as a function of mass.  The
    errors are purely from Monte Carlo statistics.  The solid line is
    our fit to the limiting value and the dotted lines its error.}
  \label{Eintfig}
\end{figure}

In the massless case, the integral is divergent in all four
triple-collinear limits.  When $i$, $j$ and $k$ are all collinear, we
have $s_{ij} \sim s_{ik} \sim s_{jk} \sim s_{ijk} \to 0$ and the leading
behaviour of the squared matrix element is $\sim1/s_{ijk}^2$.  Since the
volume of three-body phase space is $\sim s_{ijk}$, we obtain a
logarithmic divergence.  Its coefficient is the integral of the
corresponding Altarelli--Parisi splitting function (either
$P_{q\bar{q}}^{NS}$ or $P_{\bar{q}q}^{NS}$, which are equal
because of the charge-conjugation invariance of QCD).  After summing
over the four singular regions, we obtain one singular contribution for
each of the two partons in the tree-level contribution, so we
expect the coefficient of the logarithmically-enhanced term in
Eq.~(\ref{esint}) to be
\begin{equation}
  2\int_0^1 \d z P_{q {\bar q}}^{NS}(z)
  =2\left(\frac{13}4-\frac{\pi^2}2+2\zeta_3\right).
\end{equation}
That is, we expect the result retaining the quark mass to be of the form
\begin{equation}
  \label{Eint}
  \int E_S = C_F(C_F-C_A/2)\left(\frac{\as}{2\pi}\right)^2
  \left[2\left(\frac{13}4-\frac{\pi^2}2+2\zeta_3\right)\ln\frac{Q^2}{m_b^2}
  +c\right],
\end{equation}
with $c$ tending to a constant at small masses.  Our numerical results
confirm the coefficient of the log.  For the constant term we obtain the
results shown in Fig.~\ref{Eintfig}.
To obtain the limiting value, we have tried fitting various degree
polynomials in $m_b/Q$ to the points $m_b/Q\le\mu_{\mathrm{max}}$,
reducing $\mu_{\mathrm{max}}$ until the fit is acceptable.  We call the
range of values from the different fits a systematic error, which is
comparable to the statistical error, and add them in quadrature, to
give:
\begin{equation}
  c = -8.7190 \pm 0.0013.
\end{equation}

At the Z peak, logs of the bottom quark mass are not yet asymptotic.
Using the log-plus-constant approximation, we obtain
\begin{eqnarray}
  \int E_S(m_b/Q=5/91) &\approx& C_F(C_F-C_A/2)\left(\frac{\as}{2\pi}\right)^2
  \left[-0.3719\pm0.0015\right],
\end{eqnarray}
while direct integration gives
\begin{eqnarray}
  \int E_S(m_b/Q=5/91) &=& C_F(C_F-C_A/2)\left(\frac{\as}{2\pi}\right)^2
  \left[+0.8174\pm0.0001\right].
\end{eqnarray}
Even so, the difference between the two results is still an order of
magnitude smaller than $\as^2 m_b/M_Z$, the anticipated size of mass
corrections.

For charm quarks however, the log-plus-constant approximation works
quite well:
\begin{eqnarray}
  \int E_S(m_c/Q=1.5/91) &\approx& C_F(C_F-C_A/2)\left(\frac{\as}{2\pi}\right)^2
  \left[3.0922\pm0.0015\right],\\
  \int E_S(m_c/Q=1.5/91) &=& C_F(C_F-C_A/2)\left(\frac{\as}{2\pi}\right)^2
  \left[3.3527\pm0.0003\right].
\end{eqnarray}

We turn now to the integral $(2 \int E_A - \int E_S)$, which we claim is
finite in massless QCD\@.  If it is defined in the most natural way,
Eq.~(\ref{int}), the integrand is not piece-wise finite, making it
unsuitable for numerical integration.  However, we can rewrite it in a
form in which it is, proving the finiteness of the whole integral, and
allowing it to be performed numerically.

If we define $E_A(n)$ to be the $E$-term contribution that is
antisymmetric with respect to the direction $n$, then our integral
for the quark axis definition is
\begin{equation}
  \label{int}
  \int \left( E_A(p_1) + E_A(p_2) - E_S \right) \;.
\end{equation}
In each of the four triple-collinear limits $s_{ijk}\to0$, the integrand
diverges like $1/s_{ijk}^2$, again yielding a logarithmic divergence.
The coefficient of this divergence is either positive or negative,
depending on whether the collinear partons $ijk$ are $qq\bar{q}$ or
$q\bar{q}\bar{q}$.

However, using
the fact that $E_A$ is $C$-odd, we have the relation
\begin{equation}
  \int E_A(p_1) = \int E_A(p_2) = - \int E_A(p_3) = - \int E_A(p_4),
\end{equation}
which we can exploit to rewrite Eq.~(\ref{int}) as
\begin{equation}
  \int \left( \smfrac12E_A(p_1) + \smfrac12E_A(p_2)
    - \smfrac12E_A(p_3) - \smfrac12E_A(p_4) - E_S \right).
\end{equation}
In each of the four collinear limits, two of the $E_A$ terms have equal
and opposite divergences to each other and two of them have equal and
opposite divergences to $E_S$, yielding an integrable integrand with
a finite result.  We have thus
proved that Eq.~(\ref{int}) is finite.

Although this argument was formulated in terms of the $b$-quark axis
definition, it applies equally well to any infrared-safe definition,
like the thrust axis, since they must become equal in the
triple-collinear limit.

Since the integrand is everywhere integrable, we can use the same numerical
program as for the rest of the non-singlet contributions, and obtain the
results in Eqs.~(\ref{EASintb},~\ref{EASintT}).

\newpage

\section*{References}

% references
\def\ac#1#2#3{Acta Phys.\ Polon.\ #1 (19#3) #2}
\def\ap#1#2#3{Ann.\ Phys.\ (NY) #1 (19#3) #2}
\def\ar#1#2#3{Annu.\ Rev.\ Nucl.\ Part.\ Sci.\ #1 (19#3) #2}
\def\cpc#1#2#3{Computer Phys.\ Comm.\ #1 (19#3) #2}
\def\ib#1#2#3{ibid.\ #1 (19#3) #2}
\def\np#1#2#3{Nucl.\ Phys.\ B#1 (19#3) #2}
\def\pl#1#2#3{Phys.\ Lett.\ #1B (19#3) #2}
\def\pr#1#2#3{Phys.\ Rev.\ D #1 (19#3) #2}
\def\prep#1#2#3{Phys.\ Rep.\ #1 (19#3) #2}
\def\prl#1#2#3{Phys.\ Rev.\ Lett.\ #1 (19#3) #2}
\def\rmp#1#2#3{Rev.\ Mod.\ Phys.\ #1 (19#3) #2}
\def\sj#1#2#3{Sov.\ J.\ Nucl.\ Phys.\ #1 (19#3) #2}
\def\zp#1#2#3{Z.\ Phys.\ C#1 (19#3) #2}

\begin{enumerate}

\item \label{electroweak}
The LEP collaborations ALEPH, DELPHI, L3, OPAL, the LEP Electroweak
Working Group, and the SLD Heavy Flavour Group,
``A Combination of Preliminary Electroweak Measurements and Constraints
on the Standard Model'', report CERN--EP/99--15, February 1999.
%Final errors: AFB(b)=2.1%, AFB(c)=6.2%, ALRFB(b)=4.0%, ALRFB(c)=6.2%

\item \label{NLC}
K.\ M\"onig,
``Running TESLA on the Z Pole'', talk given at the Worldwide Study on
Physics and Experiments with Future Linear $e^+e^-$ Colliders, Sitges,
Spain, April 28--May 5, 1999, available from
{\tt http://www.cern.ch/Physics/LCWS99/talks.html}.
%Conclusion: 1 per mille accuracy on ALRFB(b) from 50 days running
%(systematics dominated: statistics ~ 0.4 per mille)

\item \label{JLZ}
J.\ Jers\'ak, E.\ Laermann and P.M.\ Zerwas, \pl{98}{363}{81}, \pr{25}{1218}{82}
(Erratum \pr{36}{310}{87}).

\item \label{arbuzov}
A.B.\ Arbuzov, D.Yu.\ Bardin, A.\ Leike, Mod.\ Phys.\ Lett.\ A7 (1992) 2029
(Erratum Mod.\ Phys.\ Lett.\ A9 (1994) 1515).

\item \label{DLZ}
A.\ Djouadi, B.\ Lampe and P.M.\ Zerwas, \zp{67}{123}{95}.

\item \label{lampe}
B.\ Lampe, ``A Note on QCD corrections to $A^b_{FB}$ using thrust to
determine the $b$-quark direction'', report MPI-PHT-96-14A, July 1996
(hep-ph/9812492).

\item \label{AL}
G.\ Altarelli and B.\ Lampe, \np{391}{3}{93}.

\item \label{RvN}
V.\ Ravindran and W.L.\ van Neerven, \pl{445}{214}{98}.

\item \label{BLRV}
A.\ Blondel, B.W.\ Lynn, F.M.\ Renard and C.\ Verzegnassi, \np{304}{438}{88}.

\item \label{HQ}
A.H.\ Hoang, M.\ Je\.zabek, J.H.\ K\"uhn and T.\ Teubner, \pl{338}{330}{94}.

\item \label{HQresum}
M.H.\ Seymour, \np{436}{163}{95}.

\item \label{abbaneo}
D.\ Abbaneo et al., Eur.\ Phys.\ J.\ C4 (1998) 185.

\item \label{ERT}
R.K.\ Ellis, D.A.\ Ross and A.E.\ Terrano, \np{178}{421}{81}.

\item \label{CFP}
G.\ Curci, W.\ Furmanski and R.\ Petronzio, \np{175}{27}{80};
G.\ Altarelli, \prep{81}{1}{82}; and references therein.

\item \label{Rhad2}
K.G.\ Chetyrkin, A.L.\ Kataev and F.V.\ Tkachov, \pl{85}{277}{79};
W.~Celmaster and R.J.\ Gonsalves, \prl{44}{560}{80}.

\item \label{RivNNP}
P.J.\ Rijken and W.L.\ van Neerven, \np{487}{233}{97};
\pl{386}{422}{96}.

\item \label{CS}
S.\ Catani and M.H.\ Seymour, \pl{378}{287}{96}, \np{485}{291}{97} 
(Erratum \np{510}{503}{97}).

\item \label{ZvN}
E.B.\ Zijlstra and W.L.\ van Neerven, \np{383}{525}{92}.

\item  \label{GG}
W.T.\ Giele and E.W.N.\ Glover, \pr{46}{1980}{92}.

\item \label{RivN}
P.J.\ Rijken and W.L.\ van Neerven, \pl{392}{207}{97}.

\end{enumerate}

\end{document}